\documentclass[aps,prx,showpacs,superscriptaddress,groupedaddress]{revtex4-1}  
\usepackage{graphicx}  
\usepackage{dcolumn}   
\usepackage{bm}        
\usepackage{amssymb}   
\usepackage{amsmath}   

\usepackage{color} 

\usepackage{multirow}
\usepackage{enumerate}
\usepackage{comment}
\usepackage{tabularx}
\usepackage{booktabs}

\newcommand{\irm}{{\rm i}}
\newcommand{\erm}{{\rm e}}
\newcommand{\beq}{\begin{equation}}
\newcommand{\eeq}{\end{equation}}

\begin{document}

\title{Observation of a potential future sensitivity limitation from ground motion at LIGO Hanford}

\author{J. Harms$^{1,2}$}
\author{E. L. Bonilla$^{3}$}
\author{M. W. Coughlin$^{4,5}$}
\author{J. Driggers$^{6}$}
\author{S. E. Dwyer$^{6}$}
\author{D. J. McManus$^{7}$}
\author{M. P. Ross$^{8}$}
\author{B. J. J. Slagmolen$^{7}$}
\author{K. Venkateswara$^{8}$}
\affiliation{$^{1}$Gran Sasso Science Institute (GSSI), I-67100 L'Aquila, Italy}
\affiliation{$^{2}$INFN, Laboratori Nazionali del Gran Sasso, I-67100 Assergi, Italy}
\affiliation{$^{3}$Stanford University, Stanford, California 94305, USA}
\affiliation{$^{4}$School of Physics and Astronomy, University of Minnesota, Minneapolis, Minnesota 55455, USA}
\affiliation{$^{5}$Division of Physics, Math, and Astronomy, California Institute of Technology, Pasadena, CA 91125, USA}
\affiliation{$^{6}$LIGO Hanford Observatory, Richland, WA 99352, USA}
\affiliation{$^{7}$OzGrav, Australian National University, Research School of Physics, Canberra, Australian Capital Territory 2601, Australia}
\affiliation{$^{8}$Center for Experimental Nuclear Physics and Astrophysics, University of Washington, Seattle, Washington, 98195, USA}

\date{\today}

\begin{abstract}
A first detection of terrestrial gravity noise in gravitational-wave detectors is a formidable challenge. With the help of environmental sensors, it can in principle be achieved before the noise becomes dominant by estimating correlations between environmental sensors and the detector. The main complication is to disentangle different coupling mechanisms between the environment and the detector. In this paper, we analyze the relations between physical couplings and correlations that involve ground motion and LIGO strain data h(t) recorded during its second science run in 2016 and 2017. We find that all noise correlated with ground motion was more than an order of magnitude lower than dominant low-frequency instrument noise, and the dominant coupling over part of the spectrum between ground and h(t) was residual coupling through the seismic-isolation system. We also present the most accurate gravitational coupling model so far based on a detailed analysis of data from a seismic array. Despite our best efforts, we were not able to unambiguously identify gravitational coupling in the data, but our improved models confirm previous predictions that gravitational coupling might already dominate linear ground-to-h(t) coupling over parts of the low-frequency, gravitational-wave observation band. 
\end{abstract}

\maketitle
\section{Introduction}
Understanding and improving low-frequency noise in current and future gravitational-wave (GW) detectors requires a detailed understanding of a detector's environment. Environmental noise couplings involve seismic \cite{OFW2012,MaEA2015}, acoustic \cite{EfEA2015,MaEA2016}, and electromagnetic fields \cite{AbEA2016e,CiEA2018}. When ambient fields, and also moving and vibrating objects, produce mass-density fluctuations, then direct gravitational coupling with the test masses gives rise to gravity noise in the detector, also known as Newtonian noise (NN) \cite{Sau1984,Har2019}. Newtonian noise of seismic and acoustic origin is predicted to limit sensitivity of LIGO and Virgo detectors in the frequency range 10\,Hz -- 20\,Hz once the sensitivities have progressed towards their design targets \cite{DHA2012,Har2019,FiEA2018}. 

It was suspected early on that seismic surface waves, i.e., Rayleigh waves, would be the dominant source of NN, which led to the first two detailed investigations of NN \cite{HuTh1998,BeEA1998}. This assumption was later confirmed for the LIGO detectors by analyses showing that the dominant seismic sources are local, e.g., part of the detector infrastructure such as ventilation systems and pumps, and by observing seismic-wave speeds consistent with Rayleigh waves \cite{CoEA2018a}. While Rayleigh waves also play an important role for Virgo, it is yet unclear whether they form the dominant noise contribution (acoustic fields being another contender) \cite{FiEA2018,TrEA2019}. The main analyses are still underway, but since the distance of Virgo's test masses to ground is several meters due to a basement floor, it is expected that seismic NN is significantly reduced in Virgo \cite{HaHi2014}. Also, the current sound level inside the Virgo buildings is significantly higher than in LIGO, which means that acoustic NN might well turn out to be the dominant NN contribution \cite{FiEA2018}. 

All significant environmental noises in the current detectors exist, self-evidently, because it is very hard to suppress them. For example, the sources of seismic noise, by which we mean noise from any ground and structural vibrations, can in principle be identified and either one monitors these vibrations to perform a noise cancellation (e.g., associated with laser-beam jitter \cite{DrEA2019}), or one reduces vibrations (e.g., coupling through scattered light \cite{HeEA2019}), but it remains an enormous practical challenge to identify sources and to conceive mitigation methods. Concerning NN, the most appealing approaches at an existing detector site are to (1) reduce environmental noise, which is a valid option since the dominant sources of NN, at least at the LIGO and Virgo sites, are part of its infrastructure (actions can also be taken to avoid certain forms of off-site sources), and (2) to perform a noise cancellation \cite{Cel2000}. 

The cancellation of NN requires the deployment of sensor arrays to monitor the environmental fields. Only under near ideal conditions, the required information can be obtained from a single point; see, for example, \cite{HaVe2016}. Generally, one should expect that many sensors are required \cite{CoEA2016a,BaHa2019}. Current plans for the Virgo NN-cancellation system foresee the deployment of 30 seismometers around each test mass. The sensor data are passed through a filter, e.g., a Wiener filter calculated from sensor correlations, whose output is subtracted from a detector's GW data. The great unsolved problem is how to optimally place sensors to make sure that the information required for NN cancellation is extracted from the field. First optimization results were obtained for simplified models of seismic fields \cite{CoEA2016a,BaHa2019}, but a rigorous optimization based on seismic observations does not exist yet. The scale of this effort is immense, but the sources can be easily monitored as long as they are known, and a major advantage is that the coupling to the detector data is linear, which greatly facilitates noise cancellation.

This paper presents the most comprehensive attempt yet to identify NN of seismic origin in the LIGO Hanford detector using data from its second science run (O2). It focuses on the characterization of linear noise couplings, which means that the analysis is based on correlations between sensors and the GW detector. In Section \ref{sec:cause}, we present an analytic framework to analyze couplings in simple linear systems based on correlations. We then turn to a detailed discussion of linear seismic couplings in Section \ref{sec:coupling}, which is relevant to distinguish NN from other seismic noises. In Section \ref{sec:transferground}, the main correlation measurements and noise projections are presented mostly using the ground tiltmeter. A detailed analysis of array data is presented in Section \ref{sec:aniso}, which provides a characterization of the seismic field, a validation of the array-analysis method, as well as an updated gravitational coupling model.

\section{Cause and correlation in linear systems}
\label{sec:cause}
In the following, what we mean by cause is the set of physical forces that link different observables. Even in linear systems, the relation between cause and correlation can be complicated. For example, gravity produces a coupling between ground motion and test-mass acceleration. Elastic forces produce a coupling between the suspension point of a pendulum and the suspended test mass. These couplings generally produce correlations between observations, say, of the test-mass acceleration and ground motion. However, in the presence of multiple independent causal links between observables, it is not always straight-forward to estimate the causal links from observations of correlations even if we can assume to collect all relevant information with these observations. In the following, all quantities are understood to be complex-valued Fourier amplitudes.

\begin{figure}[ht]
\centering
\vspace*{0.1cm}
\includegraphics[width=0.9\columnwidth]{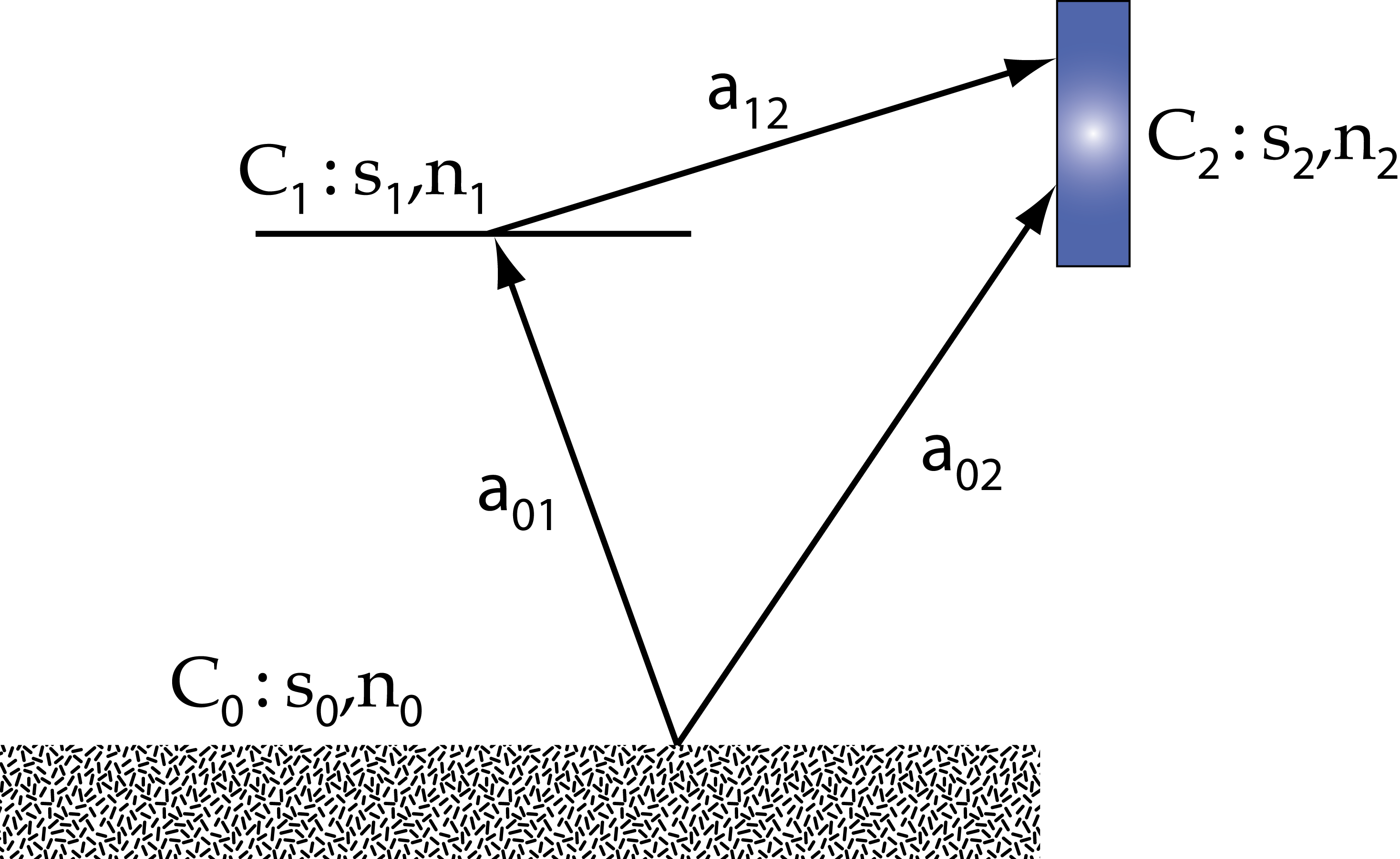}
\caption{Three-channel scheme of a directional, linear system, where $C_i$ denotes an observation, $s_i\,(n_i)$ mutually independent noise sources, which (do not) propagate through the system, and $a_{ij}$ the physical links between channels, e.g., representing elastic or gravitational forces. Directionality is a valid approximation here, since it is assumed that the physical links $a_{ij}$ are weak, and the signals obey the approximate hierarchy $a_{01}s_0\sim s_1$, and $a_{02}s_0\sim a_{12}s_1\sim s_2$.}
\label{fig:ch3}
\end{figure}

In Figure \ref{fig:ch3}, a simple linear system is depicted. It represents a three-channel system, where we understand channel $C_0$ to be an observation of the seismic field at some point, channel $C_2$ to be the acceleration of a suspended test mass, and channel $C_1$ to be an observation of vibration at some point along a mechanical link $a_{01},\,a_{12}$ between ground and test mass. In addition, we hypothesize a direct link $a_{02}$ between a point of the seismic field and test mass. This can be gravitational coupling, but in fact, applying this scheme to real-world analyses, the link $a_{02}$ incorporates anything that we do not capture with the observation of channel $C_1$. For example, there could be some charge coupling between ground and test mass (which we will rule out later), or a linear modulation of light scattered from vibrating mechanical structure circumventing $C_1$ (however, it is expected that the dominant noise from scattered light comes from up-conversion of low-frequency, high-amplitude vibrations). 

Note that the links in Figure \ref{fig:ch3} are directional and therefore violating mechanical reciprocity theorems. However, directionality is a valid approximation of the system that we are going to consider. What it means is, for example, that we assume the observed acceleration $C_2$ of the test mass to be extremely small in the relevant frequency range (displacements in the relevant frequency range are smaller than the diameter of an atomic nucleus). Therefore, the perturbation of the seismic field by an oscillating test mass through $a_{02}$ (e.g., through gravitational coupling) is negligible. In contrast, ground vibration measured at $C_0$ will be many orders of magnitude stronger than the test-mass displacement $C_2$. Concerning the vibration channel $C_1$, we assume that the mechanical links $a_{01},\,a_{12}$ form part of a seismic isolation system, and that $C_1$ experiences much smaller vibration than the ground. In other words, all physical links are assumed to be weak, and the strongest signal is entering the system in $C_0$.

The various links lead to propagation of fluctuations through the system. We consider two separate sources of fluctuations at an observation point: fluctuations $s_i$ that can propagate through the system, and fluctuations $n_i$ that cannot propagate. For example, instrumental noise of the ground sensor that observes $C_0$ will generally not propagate to other channels, and therefore is comprised in the variable $n_0$. Generally, $n_i$ can be thought of as readout noises. We will assume that correlations of fluctuations between channels vanish, i.e., $\langle n_in_j^*\rangle=0$ and $\langle s_is_j^*\rangle=0$ for $i\neq j$, so that $s_i$ can be understood as the independent fluctuations injected into the physical system at the different channels. This brings us to the following set of linear equations
\beq
\begin{split}
C_0&=s_0+n_0\\
C_1&=a_{01}s_0+s_1+n_1\\
C_2&=(a_{02}+a_{01}a_{12})s_0+a_{12}s_1+s_2+n_2
\end{split}
\eeq
The benefit from the approximate directionality of the links is that we can obtain comparatively simple relations between $a_{ij}$ and observed correlations $\langle C_iC_j^*\rangle$. This is achieved by first calculating the correlations $\langle C_iC_j^*\rangle$ between channels $i,\,j$, subsequently substituting explicit occurrences of products $\langle C_iC_j^*\rangle$ by associated complex coherences $\gamma_{ij}\equiv \langle C_iC_j^*\rangle/(\langle |C_i|^2\rangle\langle |C_j|^2\rangle)^{1/2}$ and transfer functions $\mathcal T_{ij}\equiv\langle C_i^*C_j\rangle/\langle |C_i|^2\rangle$ from channel $i$ to channel $j$ (this step is merely cosmetics), and then solving the resulting system of equations of coupling parameters $a_{ij}$ and spectral densities $<|s_i|^2>$. 

For the three-channel system in Figure \ref{fig:ch3}, the direct link $a_{02}$ is given by
\beq
a_{02}=\frac{\mathcal T_{02}(1-{\rm SNR}_1^{-2})-\mathcal T_{01}\mathcal T_{12}}{(1-{\rm SNR}_0^{-2})(1-{\rm SNR}_1^{-2})-|\gamma_{01}|^2}
\label{eq:ch3}
\eeq
where ${\rm SNR}_i^2=\langle |C_i|^2\rangle/\langle |n_i|^2\rangle\geq 1$. So not only do we need correlation measurements between channels to obtain an estimate of the link $a_{02}$, but we also need to know the spectra of the noises $n_i$ to calculate the signal-to-noise ratios (SNRs). Unfortunately, one does not always know the spectra of $n_i$, or one does not know them accurately, which poses important practical limitations to this analysis. 

According to equation (\ref{eq:ch3}), it is not enough to subtract the transfer function $\mathcal T_{01}\mathcal T_{12}$ pertaining to the isolation system from the direct transfer function $\mathcal T_{02}$ to obtain the physical link $a_{02}$. Corrections are due to the finite sensitivity of the sensor $C_i$, but even in the case of very high (infinite) SNRs, a correction needs to be applied to take into account that the estimate of the transfer function $\mathcal T_{12}$ is influenced by a possible correlation between $C_1$ and $C_2$ due to the direct link $a_{02}$ quantified by the coherence $\gamma_{01}$. If $\gamma_{01}=0$, then $a_{02}$ has no influence on the measurement of $\mathcal T_{12}$, and we get the well-known result
\beq
a_{02}=\frac{\mathcal T_{02}}{1-{\rm SNR}_0^{-2}}.
\eeq

\begin{figure}[ht]
\centering
\includegraphics[width=0.9\columnwidth]{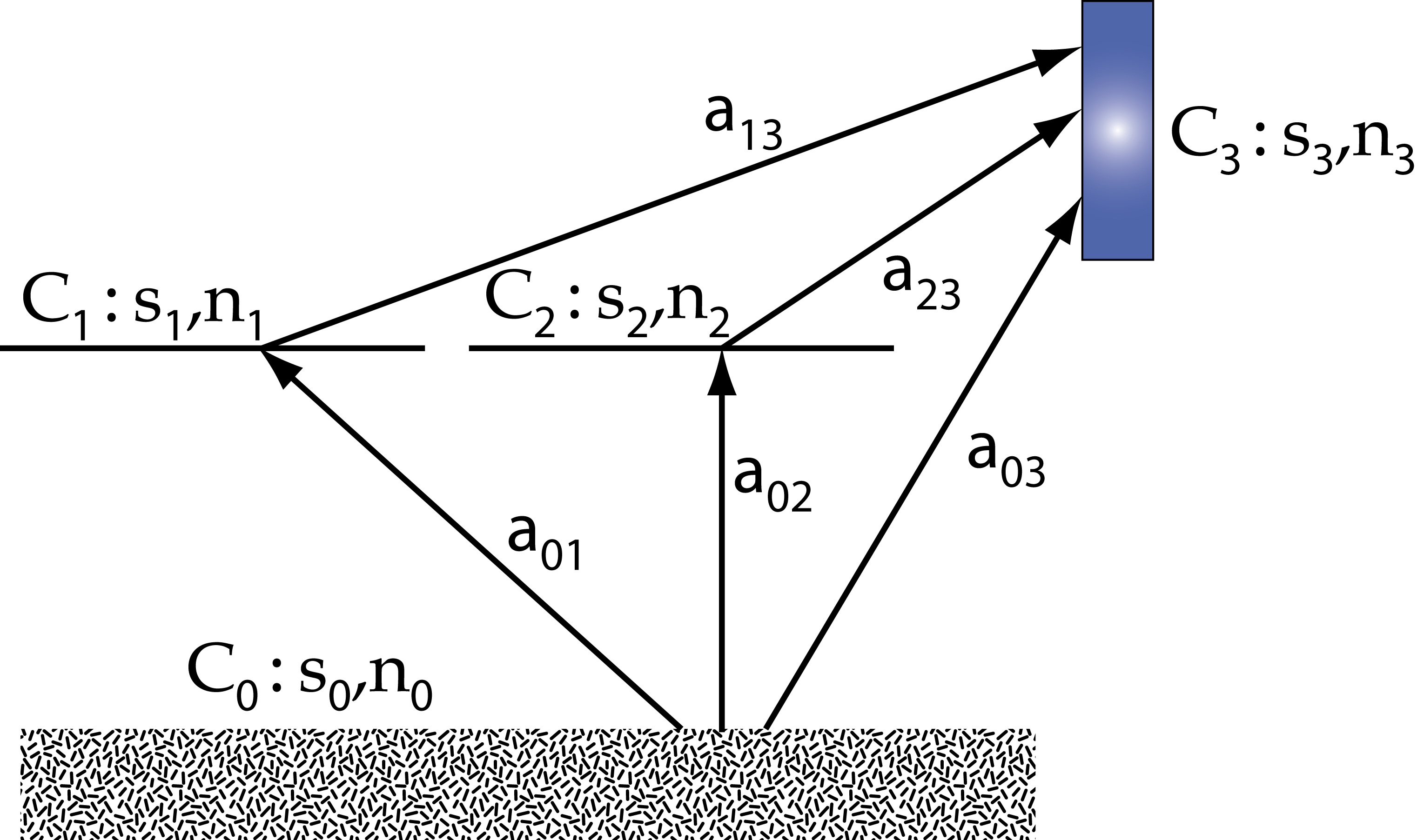}
\caption{Four-channel scheme of the directional, linear system used to study potential gravitational coupling between ground tilt $C_0$ and test-mass acceleration $C_3$.}
\label{fig:ch4}
\end{figure}

Let us now turn to the more interesting case of 2 parallel channels as depicted in Figure \ref{fig:ch4}, which forms the basis of the analysis in this paper. The solution for the link $a_{03}$ reads
\beq
\begin{split}
a_{03} &= \frac{\mathcal N_0^2\mathcal N_1\mathcal N_2-|\gamma_{01}|^2|\gamma_{02}|^2}{\mathcal N_0\left(\mathcal N_0\mathcal N_1-|\gamma_{01}|^2\right)\left(\mathcal N_0\mathcal N_2-|\gamma_{02}|^2\right)}\mathcal T_{03}\\
&\quad -\frac{\mathcal T_{01}\mathcal T_{13}}{\mathcal N_0\mathcal N_1-|\gamma_{01}|^2}-\frac{\mathcal T_{02}\mathcal T_{23}}{\mathcal N_0\mathcal N_2-|\gamma_{02}|^2},
\end{split}
\label{eq:ch4}
\eeq
where we have introduced the symbols $\mathcal N_i\equiv1-\rm SNR_i^{-2}$. It can be verified easily that this solution reduces to equation (\ref{eq:ch3}) when setting $\gamma_{02}=0$ and $\mathcal T_{02}=0$, i.e., $a_{02}=0$. 
Note that possible transverse links $a_{12},\,a_{21}$ between channels $C_1,\,C_2$ are omitted. Such a coupling can always be subsumed into the links $a_{01}$, $a_{02}$, which also means that the transverse links are superfluous to describe the dynamics of this linear system, and they cannot be estimated from correlation measurements.

\section{Coupling mechanisms between ground tilt and test-mass acceleration}
\label{sec:coupling}
Ground tilt below a test mass measured along the direction of the interferometer arm is closely linked to NN produced by seismic fields provided that the dominant contribution comes from Rayleigh waves \cite{HaVe2016}. If the seismic field is composed entirely of plane Rayleigh waves, then ground tilt is perfectly correlated with the test-mass acceleration, i.e., ground tilt can be regarded as a coherent copy of NN. In this case, a tiltmeter below the test mass could be used to completely cancel NN produced at this test mass. This also means that it is quite easy to calculate a model of the gravitational coupling as evidenced by correlations between a tiltmeter and a gravitational-wave detector. In this section, we will present the plane-wave coupling model for a tiltmeter located at the corner station of a LIGO-type detector, and compare this coupling with other proposed coupling mechanisms. 

In order to accurately represent the actual measurement that took place (see Section \ref{sec:transferground}), the coupling model needs to take into account that the tiltmeter is not necessarily located directly under the test mass, and that there is a second nearby test mass in the corner station where the same seismic field can produce NN. We present the resulting coupling between ground tilt $\tau_x$ and GW strain $h$ for Rayleigh waves with $\omega=k c(\omega)$, $c(\omega)$ being the speed of Rayleigh waves at frequency $\omega$ ($\approx 320\,$m/s at 15\,Hz), where the seismic field does not have to be isotropic, but it must be dominated by Rayleigh waves, it must be approximately homogeneous, and the waves must be approximately plane when passing the test masses:
\beq
\begin{split}
&\frac{\langle\tau_x(\omega) h^*(\omega)\rangle}{\langle|\tau_x(\omega)|^2\rangle} = \frac{2\pi G\rho_0\gamma}{L\omega^2}\\
&\quad\cdot\frac{\langle|\xi_z(\vec k)|^2k\erm^{-a k}(\cos^2(\phi)\erm^{\irm\vec k\cdot\delta\vec r_X}-\sin(\phi)\cos(\phi)\erm^{\irm\vec k\cdot\delta\vec r_Y})\rangle}{\langle|\xi_z(\vec k)|^2k^2\cos^2(\phi)\rangle},
\end{split}
\label{eq:couplingtilthoft}
\eeq
where $\vec k=k(\cos(\phi),\sin(\phi))$ is the wave vector, $a$ the height of the test mass above ground, $G$ Newton's gravitational constant, $\rho_0$ the mean density of the ground, $\gamma\approx 0.83$ a factor to take into account that density perturbations below the surface due to Rayleigh waves partially cancel the gravity perturbation from surface displacement, $L$ is the length of a LIGO interferometer arm, $\xi_z$ the vertical surface displacement, and $\delta\vec r_{\{X,Y\}}$ are the relative position vectors between tiltmeter and inner test masses of the $X$ and $Y$ arm (known as ITMX and ITMY in LIGO jargon).

\begin{figure}[ht]
\centering
\includegraphics[width=0.9\columnwidth]{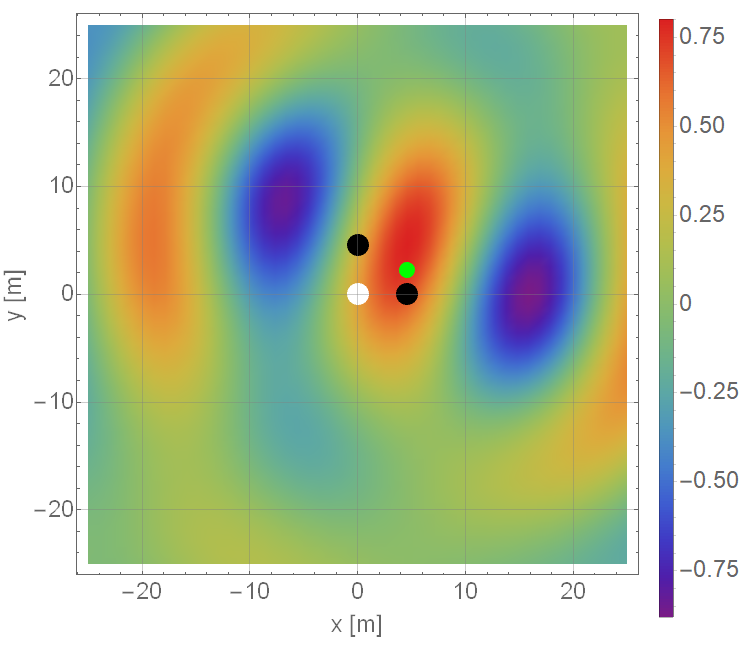}
\caption{Coupling at 15\,Hz between ground tilt of an isotropic Rayleigh field measured at position $x,\,y$ along the $x$-direction and associated NN with the positions of the two inner test masses marked in black. The white marker shows the position of the main beam splitter, and the green marker the location of the tiltmeter during the observation period used in this paper. Color scale is in arbitrary units.}
\label{fig:isotilthoft}
\end{figure}
Assuming isotropic averages in equation (\ref{eq:couplingtilthoft}), the resulting coupling at 15\,Hz is shown in Figure \ref{fig:isotilthoft} (omitting the physical unit of the coupling). It can be seen that the position the tiltmeter had during the measurements analyzed for this paper (green marker) was almost ideal to reveal a coupling with the strain data (at least when assuming isotropy of the seismic field). Of course, with a different orientation of the tiltmeter, e.g., to measure $\tau_y$ instead, the plot would look differently. Also note that, as usual, the isotropic average is real-valued even though the underlying equation represents a complex-valued coupling. 

Before turning to the actual measurements in the next section, we want to give an account of alternative coupling mechanisms between ground tilt and strain data $h(t)$, and whether we can make any ad hoc assumptions about their significance relative to gravitational coupling. These coupling mechanisms are {\it ground tilt} to:
\begin{itemize}
\item {\it rotation and translation of suspension table} to {\it test-mass displacement} via mechanical links and charge coupling between suspension frame and test mass.
\item {\it vibration of optics mounts, of suspension cage, etc} to {\it laser phase noise} via optical scattering and linear modulation.
\item {\it test-mass displacement} via gravitational coupling.
\end{itemize}
Charge coupling between ground and test mass is excluded since the vacuum chambers act as Faraday cages (there might be minor coupling of ground charges with electronics and cables). Other linear environment-to-$h(t)$ coupling observed in the past, but which is very unlikely to be captured with the deployed array, which only monitors ground motion near the test masses, include intensity modulation by vibration driven clipping such as from the elliptical baffle between the test mass and beam splitter, phase noise from Doppler shift of light produced by relative motion between optical sub-systems of the interferometer,  mode noise as in beam-size jitter from vibration in laser or angular jitter from laser-table motion \cite{DrEA2019}, vibration of electronics induced by the acoustic field modulating control currents, and vibration of cables inducing currents.

We first turn to the issue of scattered light. Scattered light is one of the major nuisances during the commissioning phase of a GW detector \cite{CaEA2013,EfEA2015}, which has often been identified as a relevant coupling mechanism between vibrations and strain data circumventing all or part of the seismic isolation system. However, models suggest that the dominant contribution to this noise comes from low-frequency ($<10$\,Hz), high-amplitude vibrations up-converted into the GW band when light is reflected from the vibrating structures \cite{OFW2012}. The argument is that a few tens of parts-per-million (ppm) light scattered from imperfect optics onto vibrating structure, experiencing a weak, i.e., \emph{linear} modulation when reflected from these structures, and then recombining partially with the laser field in the main interferometer arm should lead to negligible noise since the amount of scattered light recombining with the main field is just too small. Still, it is conceivable that strong (but linear) modulations might produce significant noise, which would then affect our analysis. 

The next coupling between ground tilt and test mass is through the suspension and seismic isolation system. Its dynamics are very complicated, and only a subset of all potentially relevant degrees of freedom are monitored \cite{MaEA2014,MaEA2015}. However, two aspects work in our favor:
\begin{itemize}
\item Any purely mechanical transmission of ground tilt to test-mass displacement must propagate through the stiff suspension platform that holds the test-mass suspension, and its displacement and rotation is monitored (albeit limited by sensitivity of the sensors).
\item Charges on the test mass might lead to additional electric coupling, e.g., between test mass and the suspension cage, which is located around and close to the test mass, circumventing the last suspension stages. The fluctuations transmitted through this coupling would still originate from motion of the suspension platform though.
\end{itemize}
Consequently, we can catch any ground-tilt to test-mass link going through the seismic isolation system by analyzing the 6 channels that monitor the 6 rotational and translational degrees of freedom of the stiff suspension platform. Any internal vibrations of the platform are relevant only at higher frequencies well above the NN band. 

\section{Observed correlations at LIGO Hanford}
\label{sec:transferground}
The system under consideration is a test mass suspended from a stiff platform whose motion can be described by 6 degrees of freedom, three translational and three rotational, assuming that internal vibration modes of this platform occur at frequencies well above 20\,Hz \cite{MaEA2015}. In the following, we will focus on two of these six degrees of freedom, the horizontal displacement L of the platform along the direction of the interferometer arm, and the pitch motion P, which is a rotation of the platform around a horizontal axis perpendicular to the interferometer arm. We verified that among all suspension-platform motions, L, P have the strongest coupling to the horizontal displacement $X_{\rm tm}$ of the test mass in the frequency band of interest. 

The data used here cover the period from Dec 1, 2016, UTC 00:00 until Feb 10, 2017, UTC 00:00. As a first step, the data were down-sampled to $f_{\rm s}=64$\,Hz and subsequently divided into non-overlapping segments of $T=8$\,s length. Next, good quality segments were selected based on the following conditions on the periodogram $|\tilde h(f)|^2/T$, where $\tilde h(f)$ is the fast Fourier transform of a segment of GW data $h(t)$ of the Hanford detector:
\begin{itemize}
\item The average over 6 frequency bins of the periodogram centered around 6\,Hz needs to lie above $4\cdot10^{-42}$\,Hz$^{-1}$
\item The periodogram at 9.3\,Hz needs to lie below $10^{-34}$\,Hz$^{-1}$
\item The average over 10 frequency bins of the periodogram centered around 20\,Hz needs to lie below $10^{-40}$\,Hz$^{-1}$
\end{itemize} 
Clearly, this selection scheme is tailored to the sensitivity of the Hanford detector at that time depending also on the width $1/T$ of a frequency bin. A histogram of periodograms showed minimal variation over time, which means that all segments represented a high-sensitivity state of the detector. Furthermore, only a negligible number of potentially useful segments were excluded from the analysis. 

\begin{figure}[ht]
\centering
\includegraphics[width=0.9\columnwidth]{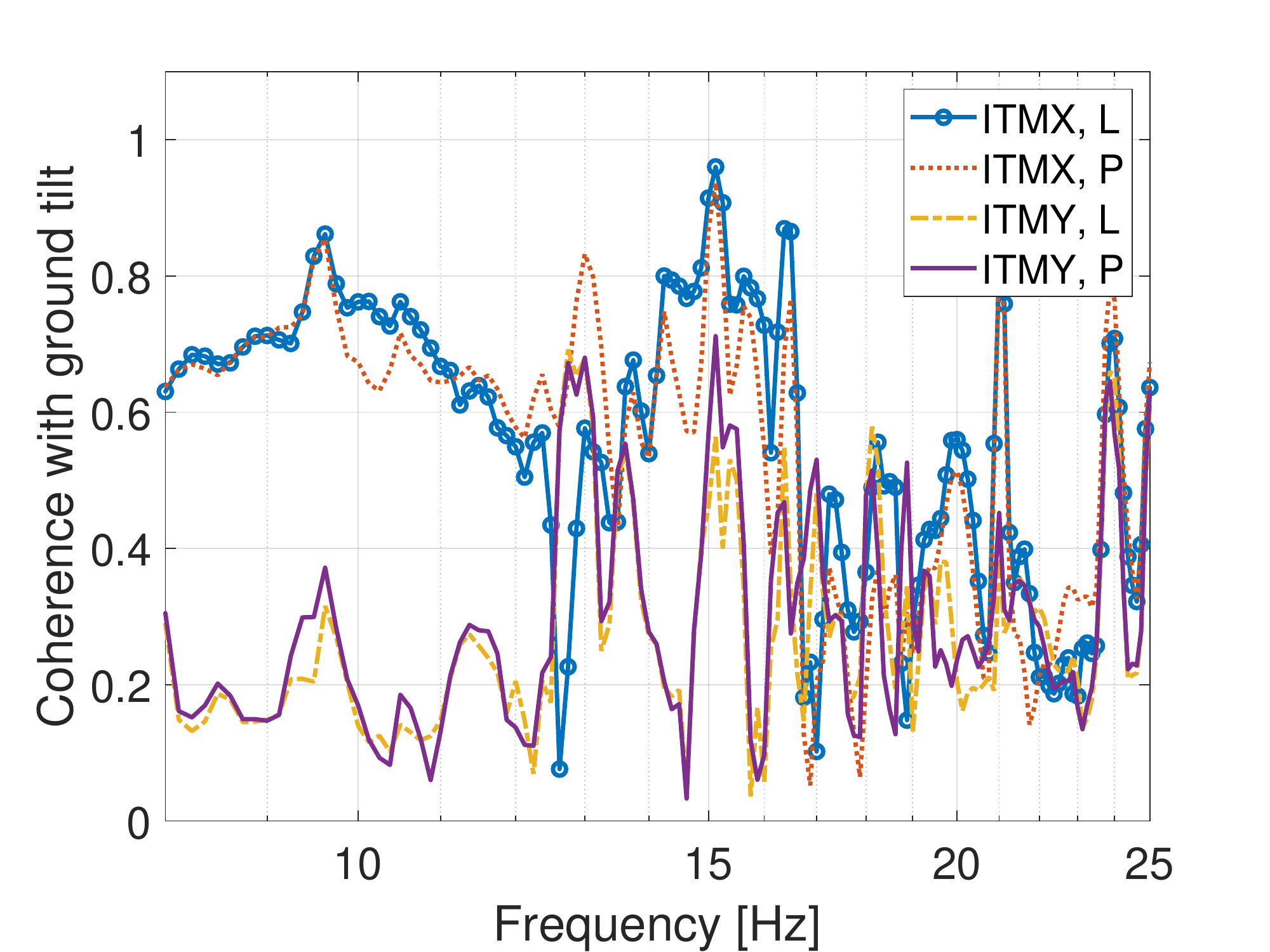}
\includegraphics[width=0.9\columnwidth]{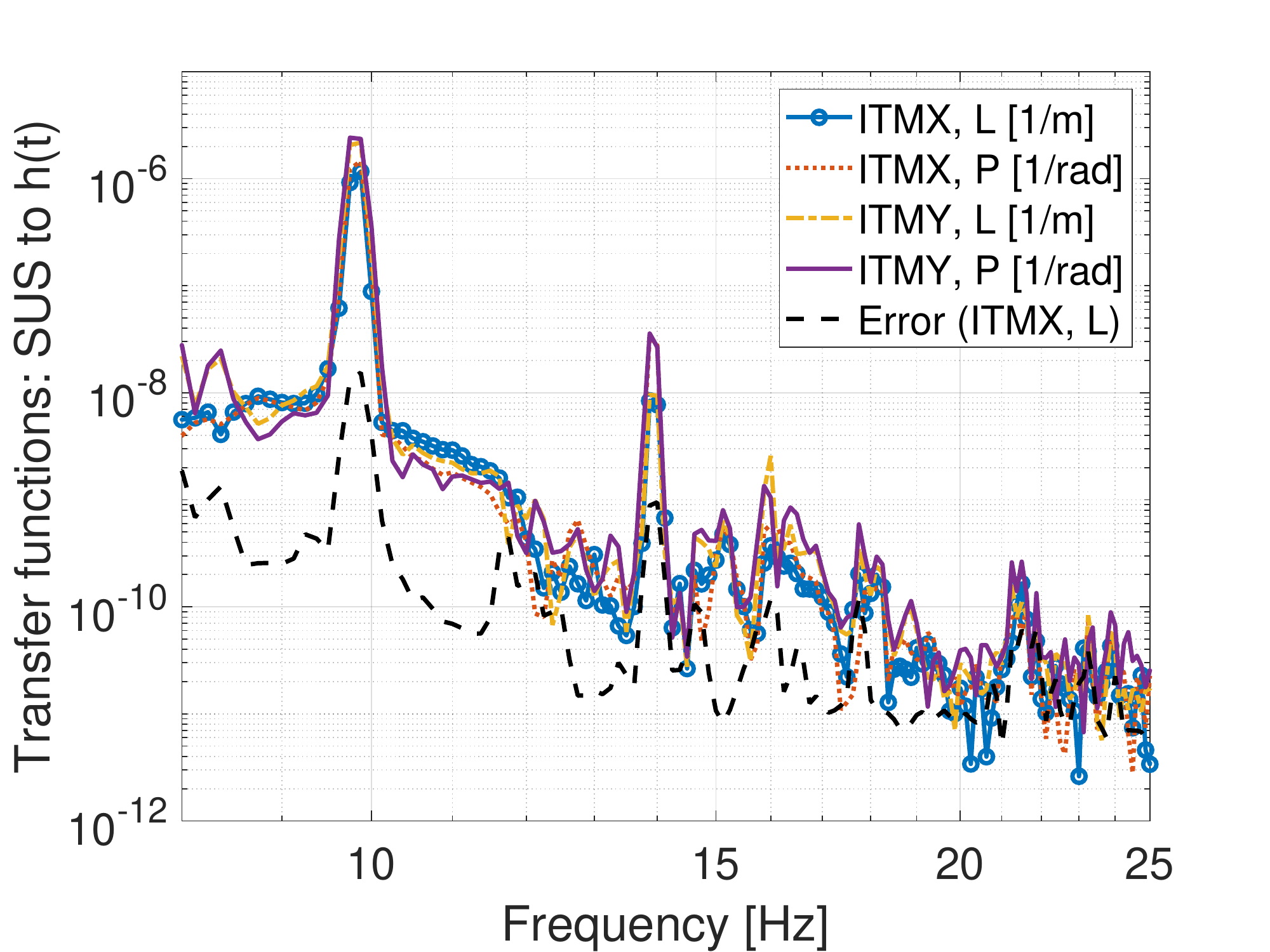}
\caption{Top: Average coherences between ground tilt and suspension-point channels. Bottom: Transfer functions from suspension point to h(t) including the statistical estimation error of the ITMX, L transfer function.}
\label{fig:sushoft}
\end{figure}

For the subsequent analysis, one needs transfer-function and coherence measurements between all channels (ground motion, L,\,P at ITMX and ITMY, and $h(t)$). Ground motion channels include L-4C vertical sensors of a seismometer array, and a tiltmeter deployed near ITMX oriented along the X-arm \cite{CoEA2018a}. Figure \ref{fig:sushoft} shows coherence measurements between ground tilt and ITMX/Y L,P (top) and transfer function measurements from ITMX/Y L,P to $h(t)$. 
It is not surprising that coherence with ITMX L, P channels is higher since the tiltmeter was located closely to ITMX. Still, coherence with ITMY L, P channels is significant at some frequencies above 12\,Hz. It is worth noting that the coherence between ground tilt and ITMX channels is very high below about 17\,Hz, which means that the active isolation during the respective observation period was relatively ineffective with respect to ground tilt in the NN band.

The transfer functions (note the different units of L and P channels) have similar values at ITMX and ITMY. Again, this is not surprising since all the degrees of freedom of the suspension platform are monitored by the same type of seismic sensors, i.e., by using linear combinations of data of these sensors to reconstruct, e.g., L and P motion, and the distance between sensors is of order meters, which means that absolute values of rotation in radians are similar to values of displacement in meters. The bottom plot also contains the estimation error of the ITMX L transfer function
\beq
\delta\mathcal T_{\rm ITMX, L}(f)= \sqrt{S_h(f)/(N S_{\rm ITMX, L}(f))},
\eeq
where $S_h$ is the spectral density of $h(t)$, $S_{\rm ITMX, L}$ is the spectral density of ITMX L, and $N$ is the number of 8\,s segments used to estimate the transfer function ($N\sim 450,000$). The length of a segment is therefore a compromise between providing sufficiently good spectral resolution and achieving sufficient averaging for the correlation measurement. The result shows that transfer-function estimates are dominated by statistical noise above 20\,Hz, which comes from the fact that the portion of $h(t)$ correlated with ITMX L is very weak compared to other low-frequency noise in the LIGO Hanford detector. 

Next, we present the spectra of noise contributing to $h(t)$ via a linear ground-to-h(t) coupling. We start with the simpler case where only the ground tilt $\tau_x$ monitored by the tiltmeter is considered. The corresponding contribution of $\tau_x(t)$ to $h(t)$ is given by
\beq
S^h_{\rm tilt}(f) = \frac{|\langle\tilde \tau_x(f)\tilde h^*(f)\rangle|^2}{\langle |\tilde \tau_x(f)|^2\rangle},
\eeq
where $\langle\cdot\rangle$ denotes a (cross) power spectral density, e.g. $\langle |\tilde \tau_x(f)|^2\rangle = S_{\tau_x}(f)$. The Gaussian estimation error of this noise projection is 
\beq
\mathcal E^h_{\rm tilt}(f) = S_h(f)/N,
\label{eq:errortilt}
\eeq
It should be noted that this error is only the leading term in a $1/N$ expansion, and it is also assumed that $S^h_{\rm tilt}(f)\ll S_h(f)$ \cite{Car1972}. As pointed out in \cite{CoEA2018a}, this Gaussian estimation error is virtually identical to the one obtained when time-series are correlated with a (sufficiently large) relative time slide.

The next case is the noise contribution considering all channels $\vec s(t)$ of the seismometer array. Here, not only correlations between $\vec s(t)$ and $h(t)$ need to be considered, but also correlations between seismometers. In fact, the best estimate of total ground-to-$h(t)$ coupling is derived from the output of a Wiener filter whose input consists of all the seismometer channels. Correspondingly, this noise contribution can be written
\beq
S^h_{\rm seis}(f) = \langle\,\vec{\tilde s}^{\,\dagger}(f)\tilde h(f)\rangle\cdot\langle\,\vec{\tilde s}(f)\circ\vec{\tilde s}^{\,\dagger}(f)\rangle^{-1}\cdot\langle\,\vec{\tilde s}(f)\tilde h^*(f)\rangle,
\eeq
where $*$ marks the complex conjugate, and $\dagger$ the complex transpose. The leading estimation error is now given by
\beq
\mathcal E^h_{\rm seis}(f) = S_h(f)\cdot n_{\rm s}/N,
\eeq
where $n_{\rm s}$ is the number of seismometer channels. Comparing this equation with equation (\ref{eq:errortilt}), one understands that it is always advantageous to use the smallest possible number of channels to estimate coherence, which includes the estimation of a Wiener filter. Especially in situations where there are constraints on the averaging time, e.g., a Wiener filter might have to be updated frequently, reducing the number of sensors can lead to important reduction of estimation errors. 
\begin{figure}[ht]
\centering
\includegraphics[width=0.9\columnwidth]{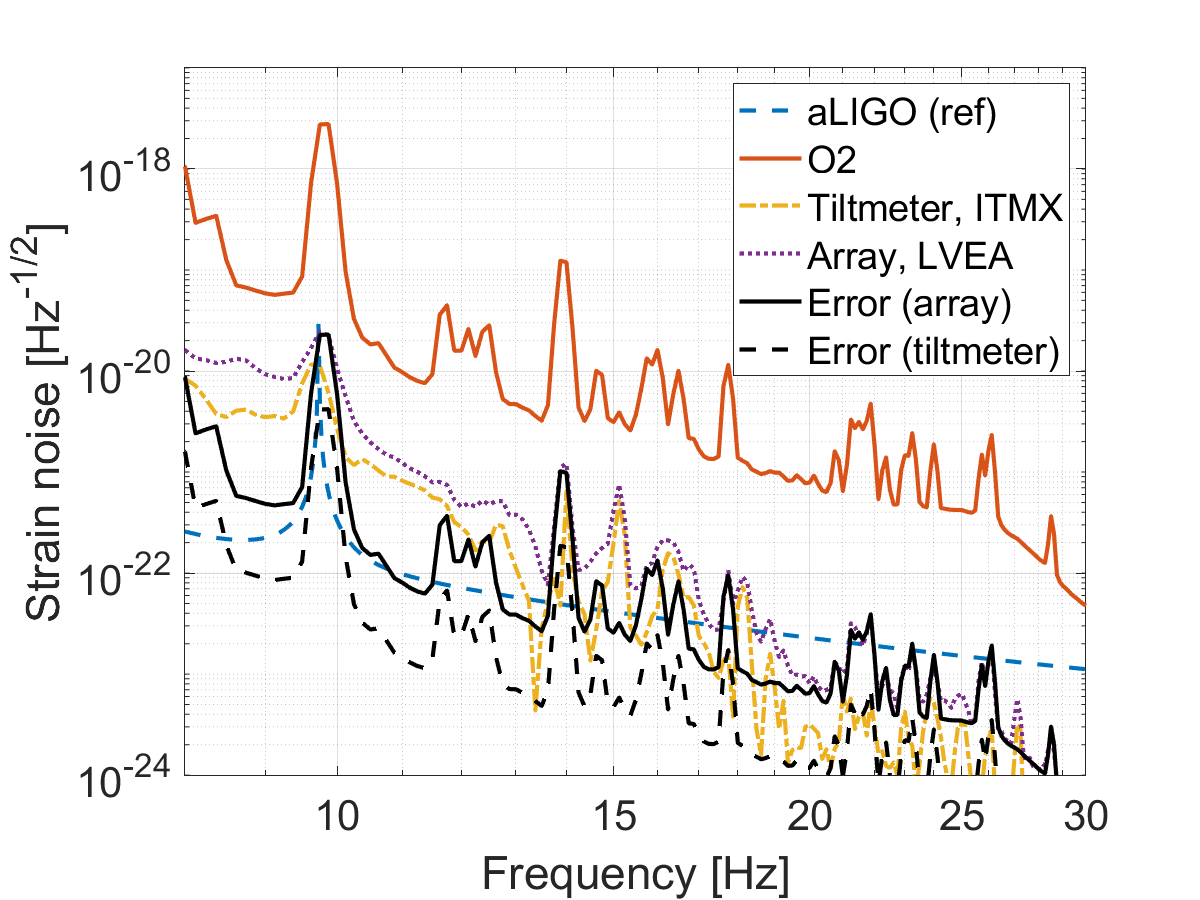}
\includegraphics[width=0.9\columnwidth]{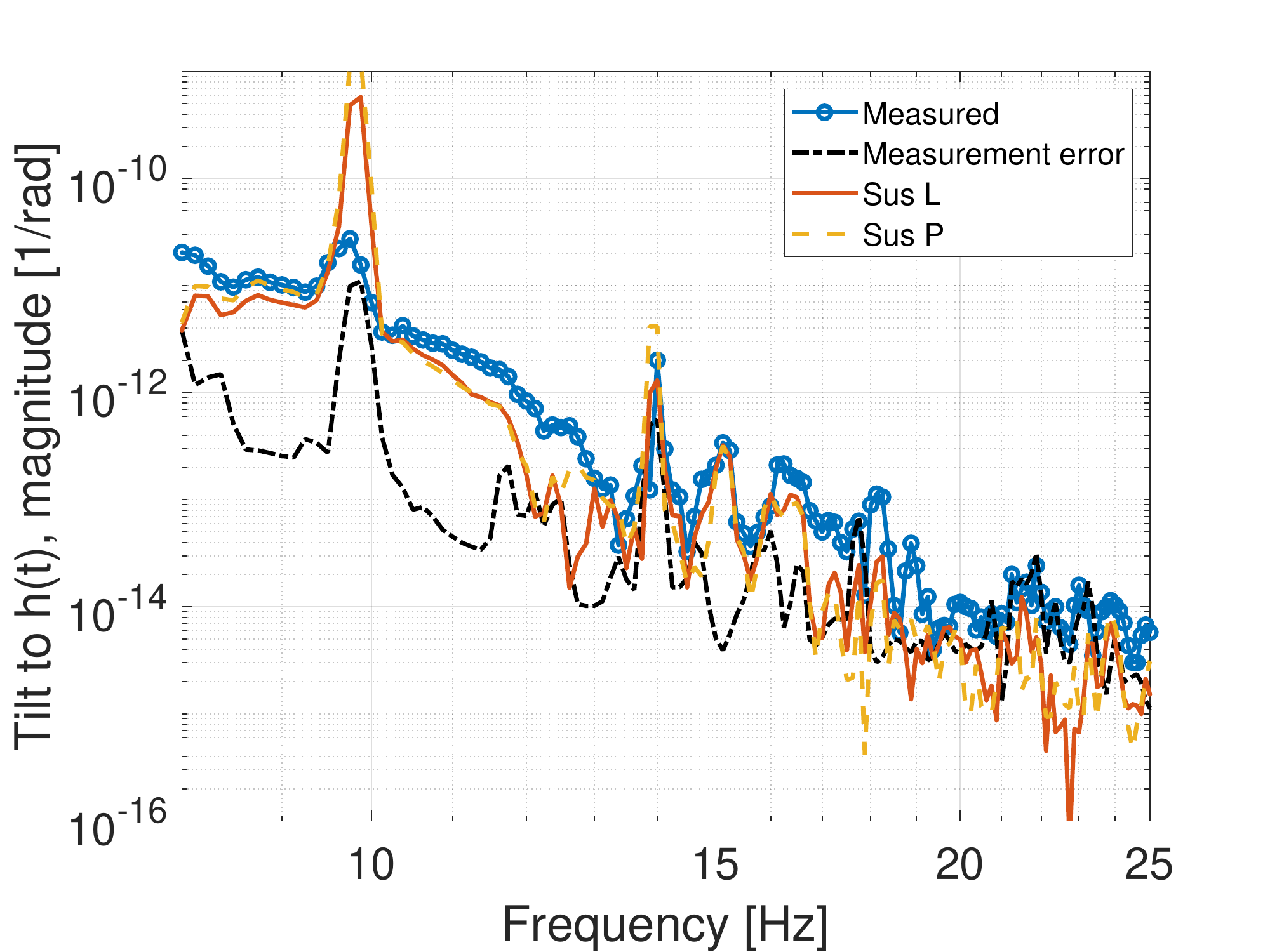}
\caption{Top: The plot shows strain noise spectra comparing the Advanced LIGO reference design sensitivity, the sensitivity reached during O2 (averaged over the days used in our analysis), and spectra of LIGO Hanford noise correlated with the tiltmeter and with the seismic array including their estimation errors. Bottom: Ground tilt-to-$h(t)$ transfer functions. Sus L, P show the multiplications of transfer functions from ground tilt to ITMX L, P and ITMX L, P to $h(t)$, while 'Measured' shows the direct ground tilt-to-$h(t)$ transfer function (together with its statistical estimation error).}
\label{fig:strainspec}
\end{figure}

The results of the noise projection are shown in Figure \ref{fig:strainspec} (top) together with an estimate of ground tilt-to-$h(t)$ transfer functions (bottom). As expected, the array recovers more correlated noise in $h(t)$ than the tiltmeter since the tiltmeter was located close to ITMX, and therefore its correlation with $h(t)$ is dominated by coupling through ITMX, while the array captures couplings equally through ITMX and ITMY. Furthermore, it might be that some of the ground-to-$h(t)$ coupling is not related to ground tilt. The estimation error for the array noise projection is a factor $\sqrt{30}\sim 5.4$ higher than for the tiltmeter. Consistent with results presented in \cite{CoEA2018a}, estimation errors dominate the two noise projections above 20\,Hz, and in addition, the influence of strong noise peaks in $h(t)$, e.g., near 10\,Hz, 14\,Hz, 18\,Hz, cannot be reduced sufficiently by averaging to obtain noise projections at these frequencies. 

The transfer functions are an updated estimate of results shown in \cite{CoEA2018a}. An update is necessary since more detailed assessments of data quality revealed the presence of outliers in the ITMX/Y L, P channels (artifacts of the data processing), which had to be removed. As a consequence, correlations involving suspension-platform channels increased. It should be stressed that this result does not directly mean that coupling of ground tilt through the suspension system explains the observed correlated noise in $h(t)$ since none of the multi-link corrections discussed in Section \ref{sec:cause} were applied here. This is being done in the next section.

\section{Estimation of the gravitational coupling}
\label{sec:aniso}
In this section, we present our final results concerning a possible direct gravitational coupling between ground tilt and $h(t)$. This requires first of all a more detailed characterization of the seismic field to validate the coupling models. Most importantly, we need to assess the degree of anisotropy and the dominant type of seismic waves.

The array analysis can be done with a standard f-k (frequency, wave vector) analysis based on the matrix $\mathcal C(\omega;\vec r_i,\,\vec r_j\,)$ of cross-spectral densities of all $N\times N$ seismometer pairs:
\beq
p(\omega,\vec k\,)=\sum\limits_{i,j=1}^N\mathcal C(\omega;\vec r_i,\,\vec r_j\,)\erm^{-\irm\vec k\cdot(\vec r_j-\vec r_i)}
\eeq
This quantity is closely related to a discrete, spatial Fourier transform of the two-point spatial correlation, but note that this expression does not take into account the unequal spacing between seismometers, which would introduce weights $w_{ij}$ in the sum. Since the matrix $\mathcal C$ is hermitian, $p(\omega,\vec k\,)$ is real valued. Limitations of this analysis mainly come from spatial aliasing, and since it is desired to sample $p(\omega,\vec k\,)$ frequently during the day to study temporal variations limiting the correlation time to evaluate the matrix $\mathcal C$, also seismometer instrumental noise plays an important role. 

\begin{figure}[ht]
\centering
\includegraphics[width=0.9\columnwidth]{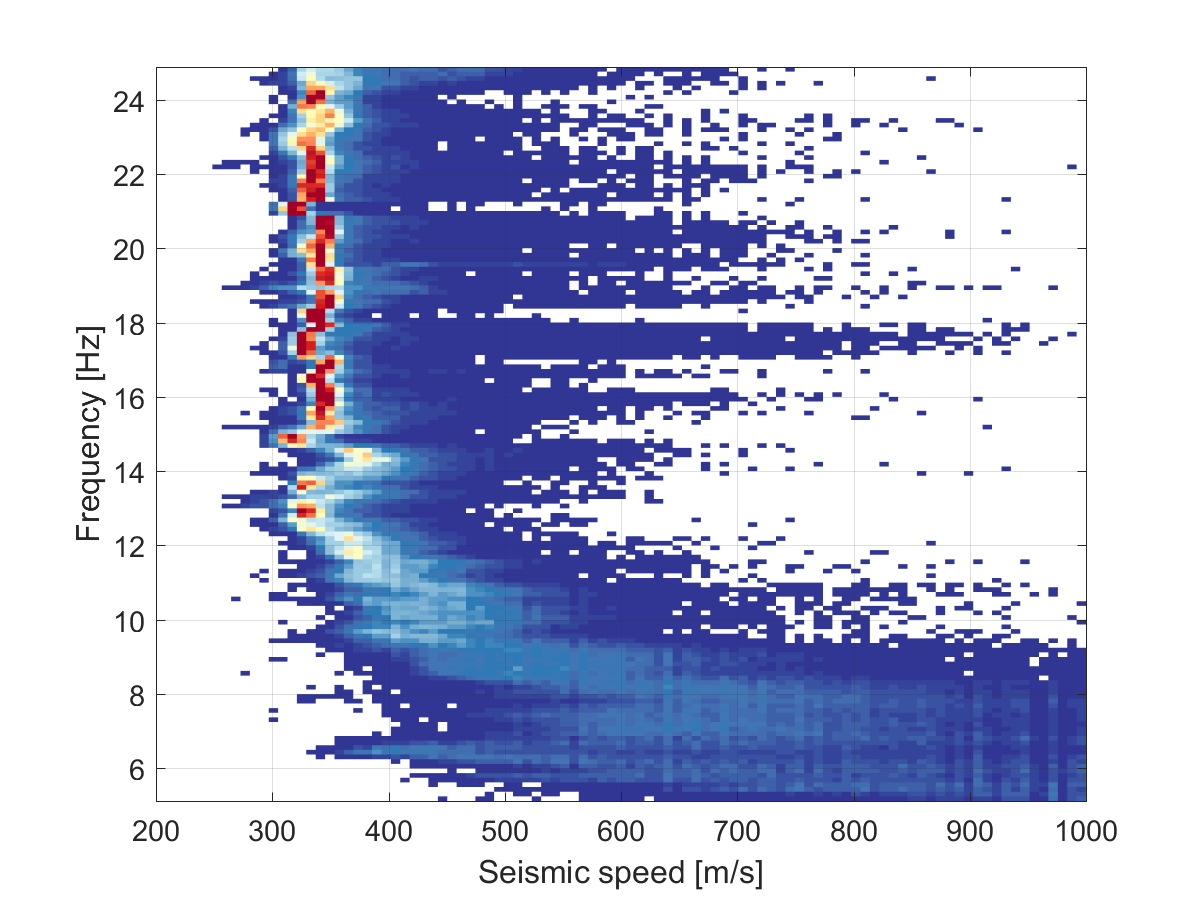}
\includegraphics[width=0.9\columnwidth]{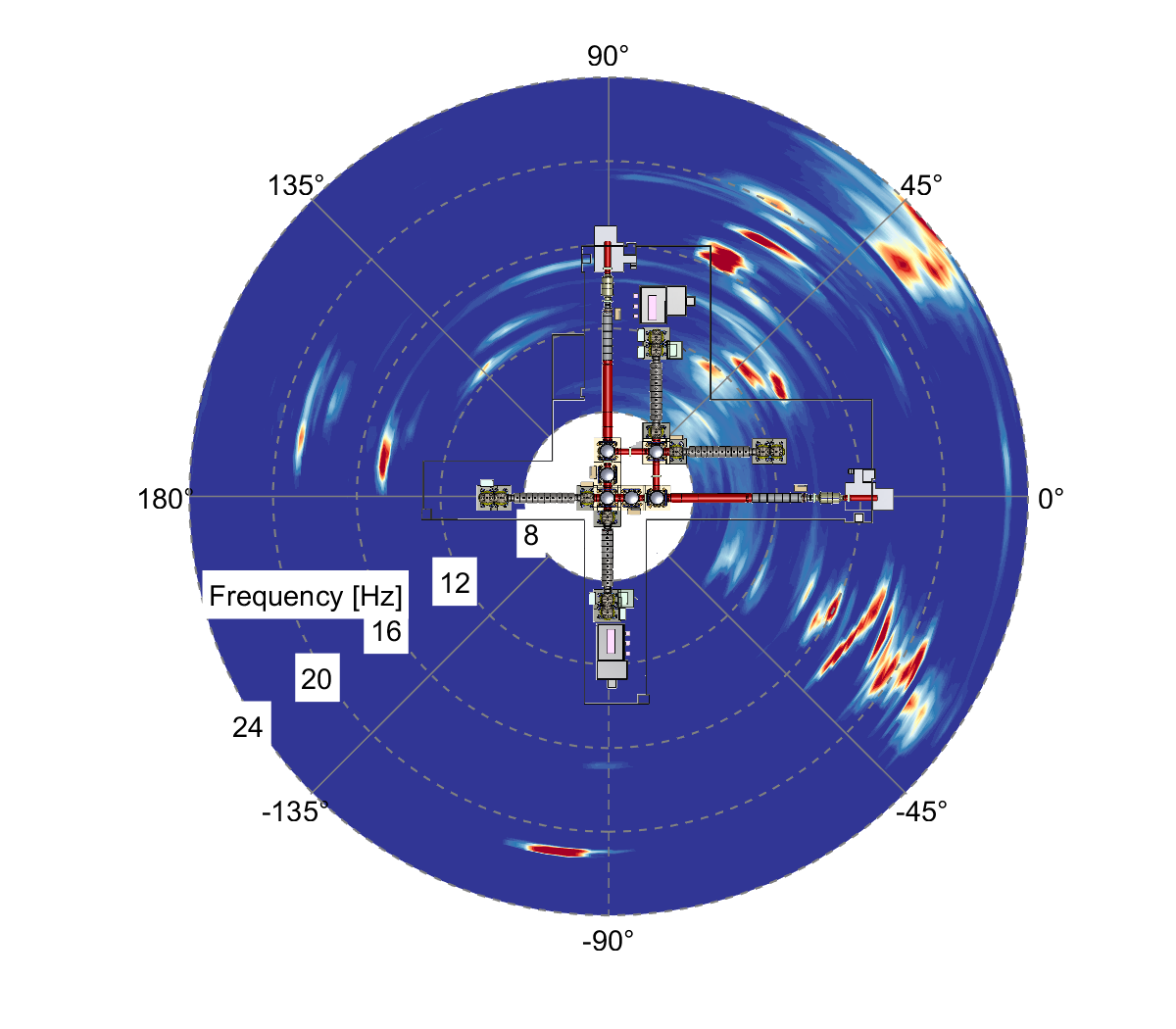}
\caption{Histograms of seismic speeds and propagation azimuths. 
Each sample in this plot is an average over 800\,s (using 8\,s Fourier transforms), and a total of about 4500 samples contribute to the histogram at each frequency.}
\label{fig:histos}
\end{figure}
For each sample of $p(\omega,\vec k\,)$, we collect the wave vector $\vec k_0=k_0(\cos(\phi),\,\sin(\phi))$ belonging to its maximum, which provides us a propagation direction $\phi$ with respect to the direction of LIGO Hanford's X-arm, and a seismic speed $c_0=\omega/k_0$. One sample is calculated every 800\,s using an average over hundred 8\,s segments for individual Fourier transforms. These values are collected over months of data and over a range of frequencies forming the histograms that can be seen in Figure \ref{fig:histos}. 

We see that  distributions of speeds and azimuths are wider below 12\,Hz, which reflects a decrease of the array's resolution in wave-vector space towards low frequencies. The most probable speeds lie around 330\,m/s between 12\,Hz and 25\,Hz, and there is a clear signature of normal dispersion of Rayleigh waves between 5\,Hz and 12\,Hz. This dispersion is explained by the Rayleigh waves' deeper sampling of the soil at larger wave lengths. Note that these long-term studies do not confirm initial indications of anomalous dispersion in the NN band, which was originally attributed to the concrete slab \cite{CoEA2018a}. This means that the slab is too thin to significantly affect waves below 25\,Hz. Also, since the wave speed coincides with the speed of acoustic waves, one might wonder if acoustic disturbances dominated the signals of seismometers. However, this can be excluded since microphone measurements show that the acoustic field has a much smaller coherence then observed by the seismic array.

The first important conclusion is that the dominant wave type is the Rayleigh wave since body waves are necessarily faster than 330\,m/s (especially considering that the array measures apparent horizontal speed of body waves). Love waves, which can propagate in shallow layers, do not contribute to vertical ground motion (or ground tilt). The second important result is that the field is highly anisotropic. In fact, most sources of the dominant Rayleigh waves lie only along 3 different directions with two additional narrowband sources around 15\,Hz and 21\,Hz. One cannot expect isotropic coupling models to be accurate.

The speed and azimuth histograms can also be used to evaluate the averages in equation (\ref{eq:couplingtilthoft}). Note that since the histograms are constructed from the maxima of $p(\omega,\vec k\,)$, our method is only an approximation to a full wave-vector based averaging, which would include additional contributions from other local maxima of $p(\omega,\vec k\,)$, i.e., representing other sub-dominant seismic waves or noise. The reason why we focus on the global maximum is that one would otherwise integrate mostly over noise instead of the interesting physical components of the wave field, and spatial resolution limits would cause a frequency-dependent bias of the results, which is difficult to understand. More sophisticated array-analysis methods such as MUSIC \cite{KrVi1996} can in principle be used to suppress contributions from noise and to overcome certain resolution limits, but since spatial spectra in the NN band typically show a single dominant mode, we decided to use a simplified approach focusing on the dominant mode.
\begin{figure}[ht]
\centering
\includegraphics[width=0.9\columnwidth]{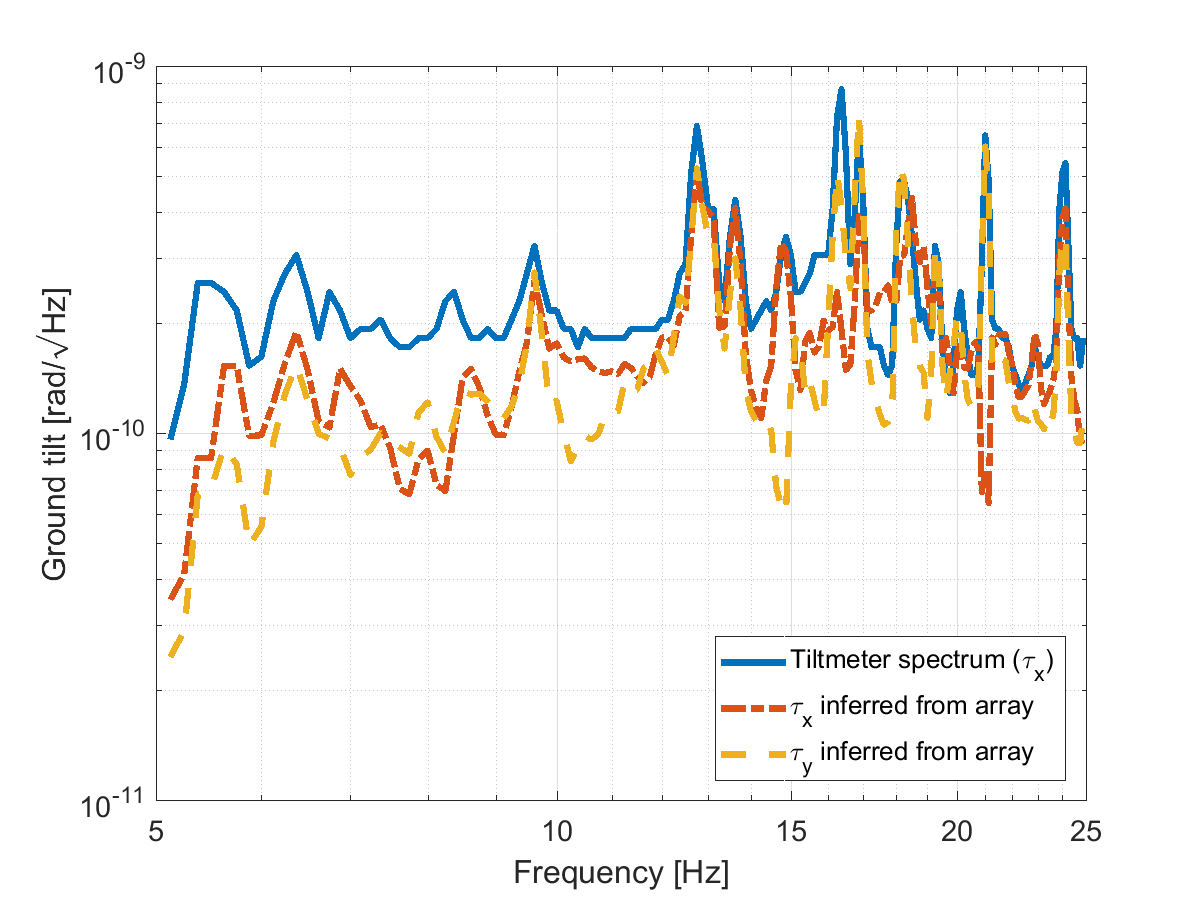}
\caption{Spectrum of the tiltmeter in comparison with the tilt spectrum inferred from the seismic-array data using equation (\ref{eq:tilt}).}
\label{fig:tiltspec}
\end{figure}

One important verification of this method is to compare tilt spectra obtained directly from the tiltmeter and inferred from the array data,
\beq
\langle|\tau_x(\omega)|^2\rangle=\langle k^2\cos^2(\phi)|\xi_z(\omega,\vec k\,)|^2\rangle,
\label{eq:tilt}
\eeq
where the left-hand side is the spectrum observed by the tiltmeter, and the right-hand side is the tilt spectrum inferred from the array data. In order to evaluate the array-inferred tilt, one needs a dispersion curve to obtain $k$ together with propagation directions to evaluate $\phi$, and the spectrum of vertical ground displacement. 

The result is shown in figure \ref{fig:tiltspec}. Especially at 10\,Hz and above, array-inferred tilt spectrum $\tau_x$ and the tiltmeter spectrum match well at most frequencies. Curiously, it is the array-inferred tilt $\tau_y$, which fits the tiltmeter spectrum at some peaks of the spectrum (16.3\,Hz and 21\,Hz). This might indicate some limitation of the array analysis, or maybe the tiltmeter is not only susceptible to $\tau_x$. It can also be that some of the approximations underlying the coupling model (only plane Rayleigh waves) are not valid at some frequencies. Below 10\,Hz, the array-inferred tilt underestimates the true ground tilt, which can be explained by the fact that the speed estimates at these frequencies are not accurate (see figure \ref{fig:histos}) and according to the tilt spectra are likely too high. 

\begin{figure}[ht]
\centering
\includegraphics[width=0.9\columnwidth]{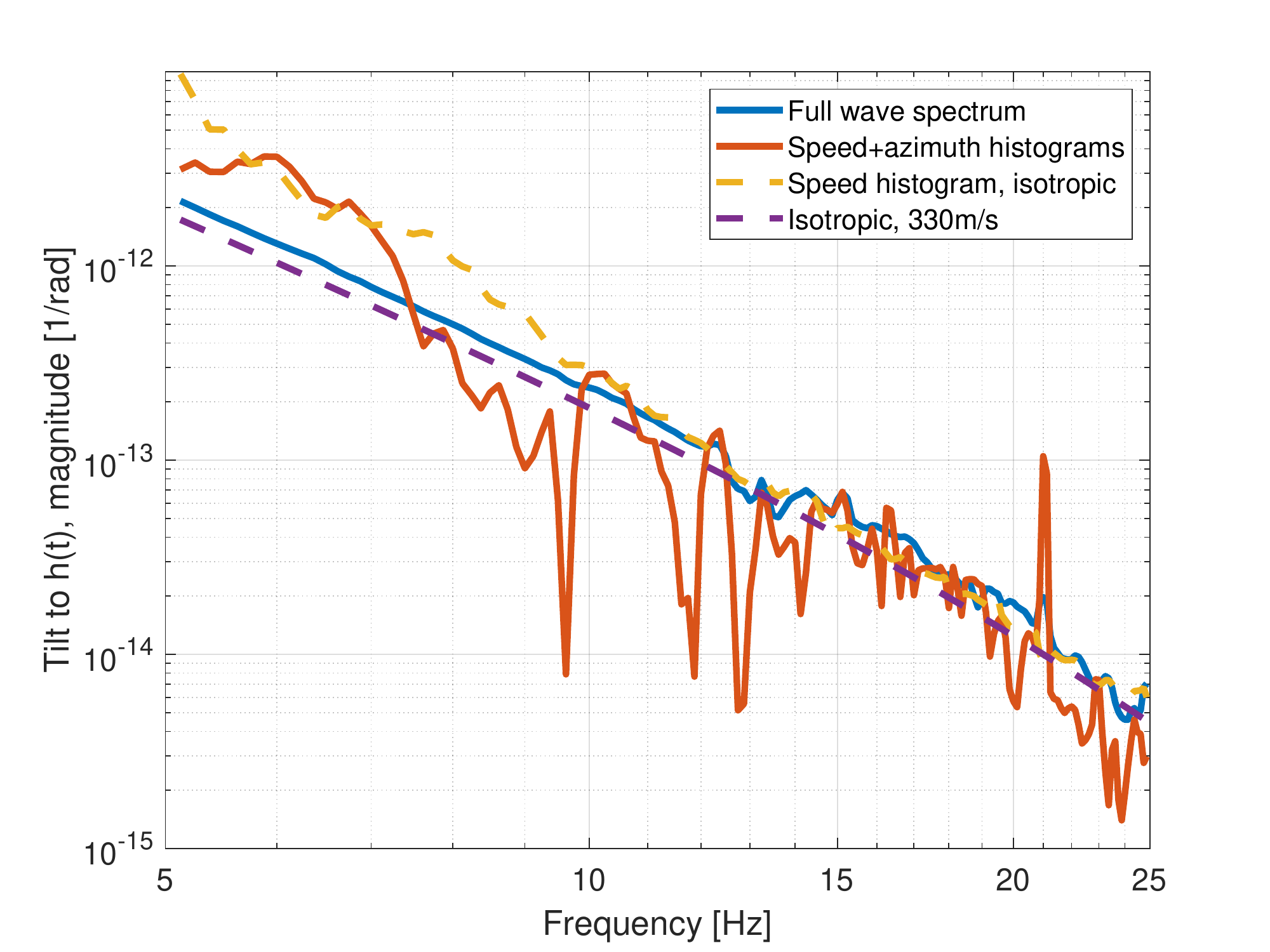}
\caption{The plot shows the transfer functions between tiltmeter and $h(t)$ using different levels of simplification of a gravitational-coupling model for Rayleigh waves propagating along a flat surface. The two dashed curves are based on ad hoc assumptions about isotropy (yellow), and isotropy and speed (violet). The two solid curves are entirely based on array data: the full wave-vector space is integrated for the blue curve, while only the discrete information of dominant modes is used for the red curve.}
\label{fig:couplings}
\end{figure}
Nonetheless, the match is good enough at many frequencies to gain trust in the array analysis, so that we can proceed with a calculation of the gravitational coupling model according to equation \eqref{eq:couplingtilthoft}. Figure \ref{fig:couplings} shows the coupling model using a varying degree of simplifying assumptions. The 'Isotropic, 330\,m/s' coupling assumes that Rayleigh waves have the same speed between 5\,Hz and 25\,Hz, and the field is isotropic. The next step is to use the speed histogram, which leads to a significant change especially below 10\,Hz where the speeds are higher. Using the speed and azimuth histograms, i.e., still only considering the maxima of $p(\omega,\,\vec k\,)$, the coupling now becomes strongly frequency dependent. Note that in anisotropic fields, the coupling is not bounded towards low or high values. For example, if at some frequency Rayleigh waves pass the tiltmeter (almost) perpendicularly to its axis, they would produce a very small tilt signal, but experience a comparatively strong gravitational coupling through ITMY. In such a seismic field, the ground tilt-to-$h(t)$ transfer function would be very high. Conversely, if at some frequency Rayleigh waves propagate at 45$^\circ$ to the X and Y arms, then very little NN would be produced (the main part experiences strong common-mode rejection), but the tiltmeter signal would be comparatively high. As a consequence, the transfer function would have a very small value at this frequency. It is therefore not surprising that the variations with frequency introduced by the anisotropic field are this strong. In fact, its form is consistent with the azimuth histogram in figure \ref{fig:histos}. The last step is to average over entire f-k maps instead of just picking their maxima. Below 10\,Hz it can be seen how the spatial resolution limit of the array causes the result to lose its dependence on anisotropies of the field since the averaging is done mostly over energy leaked from the dominant waves into neighboring wave vectors. This effect is likely significant above 10\,Hz as well, but we cannot rule out that sub-dominant waves also contribute to a change of results. 
\begin{figure}[ht]
\centering
\includegraphics[width=0.9\columnwidth]{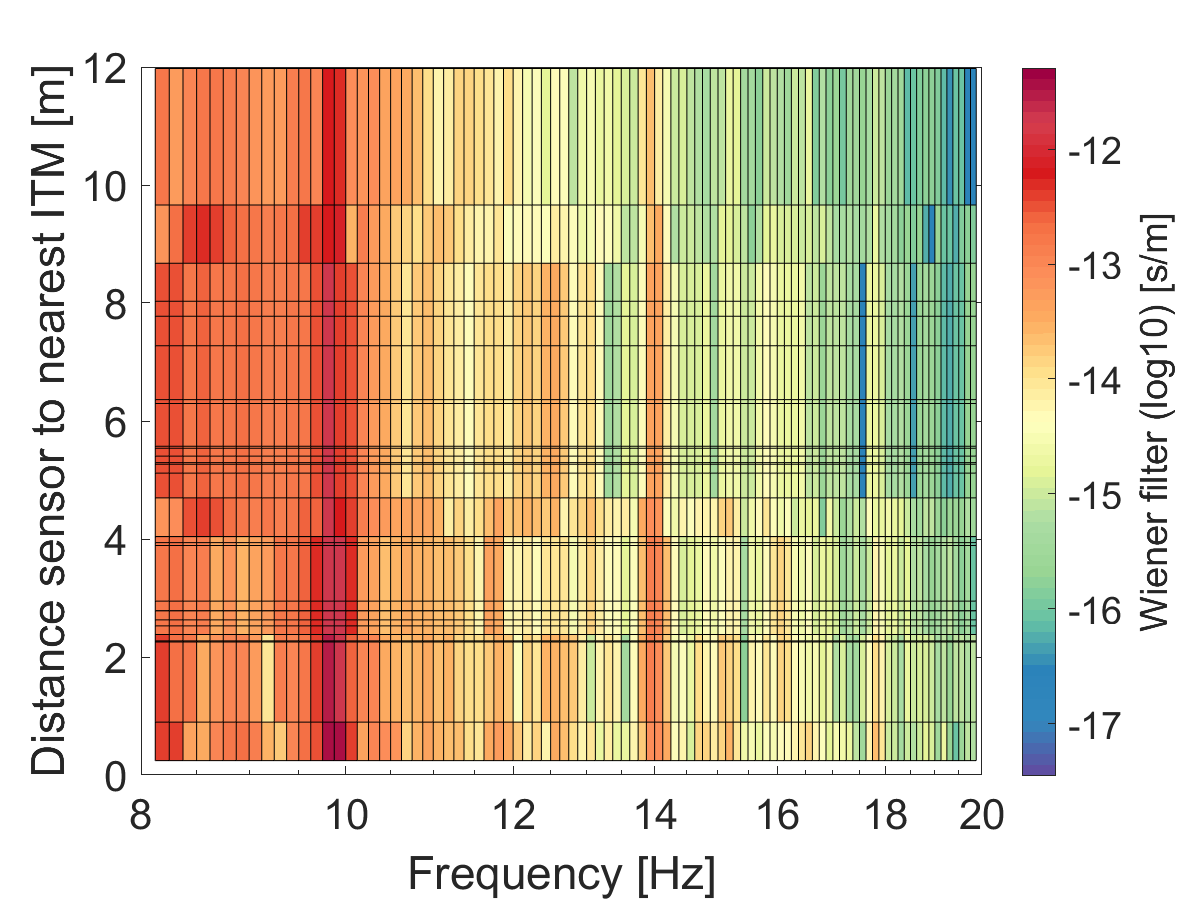}
\includegraphics[width=0.9\columnwidth]{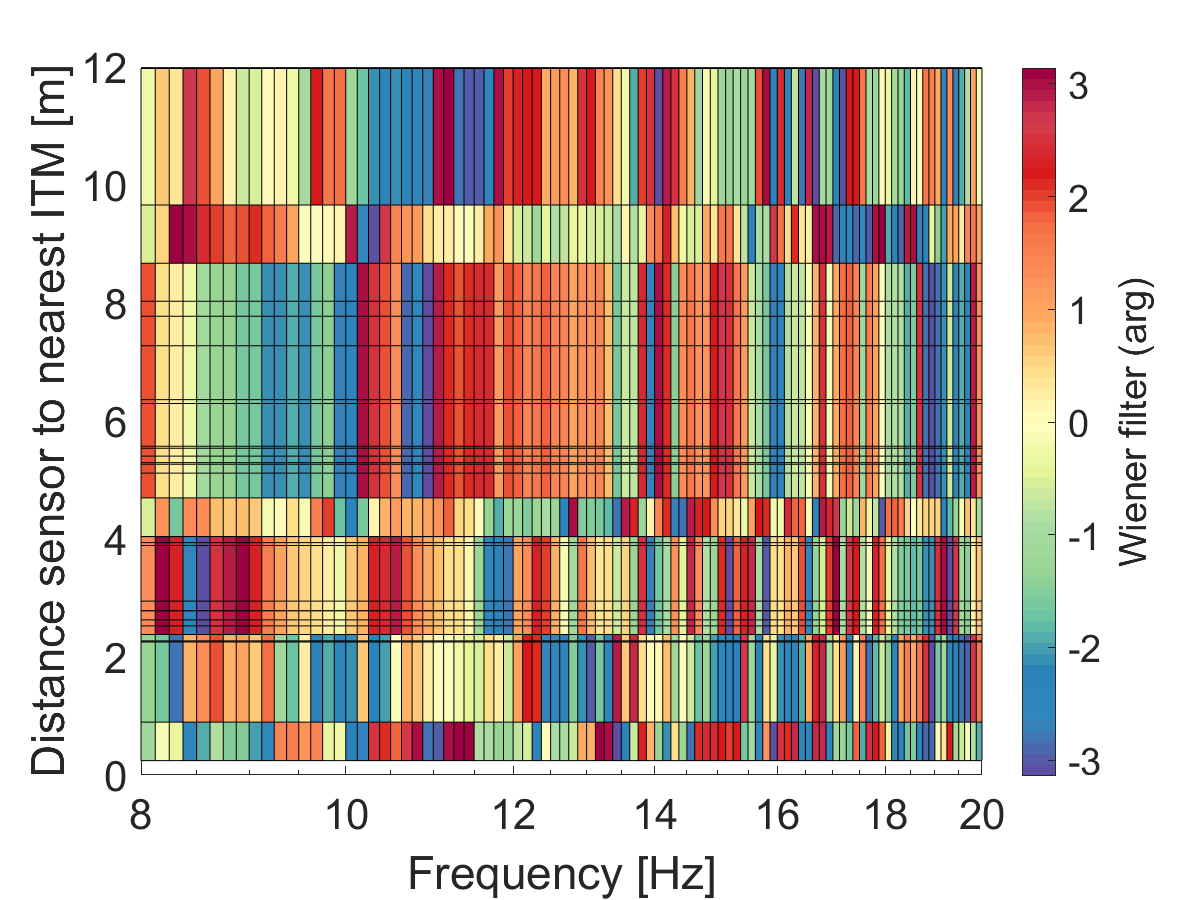}
\caption{Amplitude (top) and phase (bottom) of the array Wiener filter. The seismometers are ordered according to their distance to the nearest test mass. Seismometer distances are marked by horizontal lines, and their corresponding Wiener filter is represented by the colors below that line.}
\label{fig:wiener}
\end{figure}

A potential way to identify NN in correlation data is to study the array Wiener filter used for the noise projection in figure \ref{fig:strainspec}. Variations of phases and amplitudes of the filter over frequencies and seismometers might show inconsistencies with alternative coupling models. 

The plots in Figure \ref{fig:wiener} show the amplitude (top) and phase (bottom) of the array Wiener filter where the seismic data have units speed (measuring the time derivative of surface displacement). The seismometers are ordered according to their distance to the nearest test mass (ITMX and ITMY). The exact distance of each seismometer is indicated by a horizontal line. Its Wiener filter corresponds to the colors below that line (towards smaller distances). The first remarkable observation is that seismometers are grouped such that their Wiener filters are virtually identical in amplitude and phase, e.g., sensors between 4.7\,m and 8.5\,m to the nearest test mass. This means that none of these seismometers provides information about the seismic field that is not already contained in other seismometers of that group. The effect is to create a super-sensor with increased sensitivity via noise averaging. In total, only 7 distinct sets of filter coefficients are formed by the Wiener filter; 5 of them associated with super-sensors.

Interpreting these results is challenging, and we will explain that the anisotropy of the field as shown in Figure \ref{fig:histos} makes it practically impossible to distinguish gravitational coupling from any other coupling. Generally, the hope is that gravitational coupling makes unique predictions about the Wiener filters. This is certainly true for isotropic fields where, at LHO, the Wiener filter of sensors within about 3\,m to the test mass must have small amplitudes, increasing with distance up to some point (depending on seismometer SNR and number of seismometers) and then decreasing again towards greater distances \cite{Har2019}. A consequence is that optimal placement of seismometers in isotropic fields for NN cancellation does not include any sensors close to the test mass. However, this neglects the presence of a second test mass.

The situation is different though for anisotropic fields. The Wiener filters depend on the source distribution and, as usual, on the seismometer SNR. For example, if at some frequency there is a single seismic source, then the seismometers closest to this source are predicted to have the smallest filter amplitudes and the more distant ones higher amplitudes. In this way, the Wiener filter compensates for the decrease in seismic amplitude with distance to the source, but this also depends on seismometer SNR, which limits the amplitudes of lower-SNR sensors, which can also lead to the creation of super-sensors to overcome sensitivity limitations. In contrast to the isotropic field, there is in any case no loss of correlation with $h(t)$ for seismometers close to the test mass. The fact that gravitational coupling in anisotropic fields does not predict loss of correlation at small distances makes it hard to distinguish it from other couplings that enter locally via transmission of vibrations through the suspension systems or by scattered light. A deeper analysis using models of all the coupling mechanisms might be able to discriminate between them based on the Wiener filters, but such an effort is beyond current modeling capabilities. In conclusion, the Wiener filters, while containing interesting structures and information, do not help us with the identification of NN in LIGO Hanford mostly because of the complexity of the seismic field and its strong anisotropy.

\begin{figure}[ht]
\centering
\includegraphics[width=0.9\columnwidth]{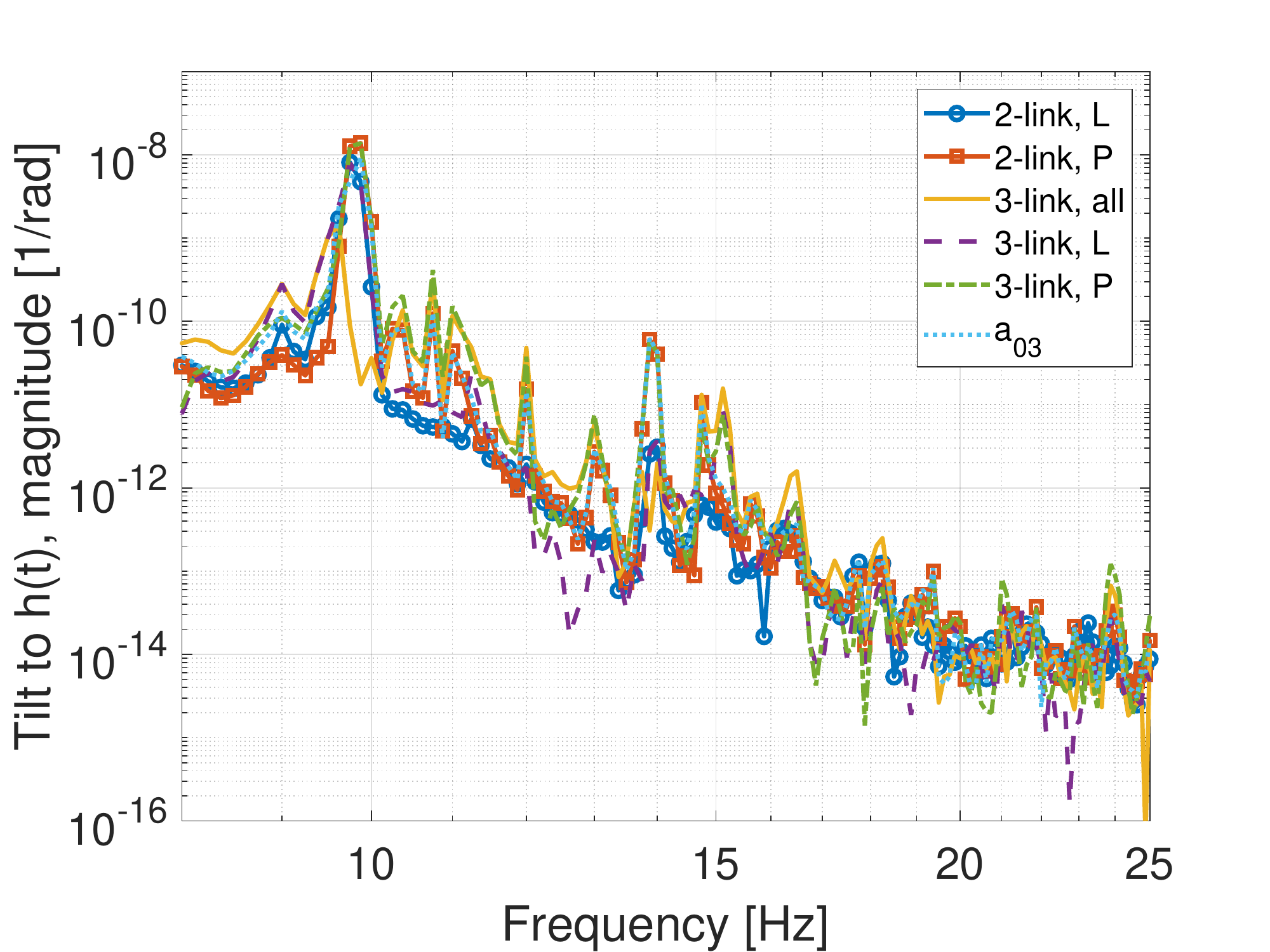}
\includegraphics[width=0.9\columnwidth]{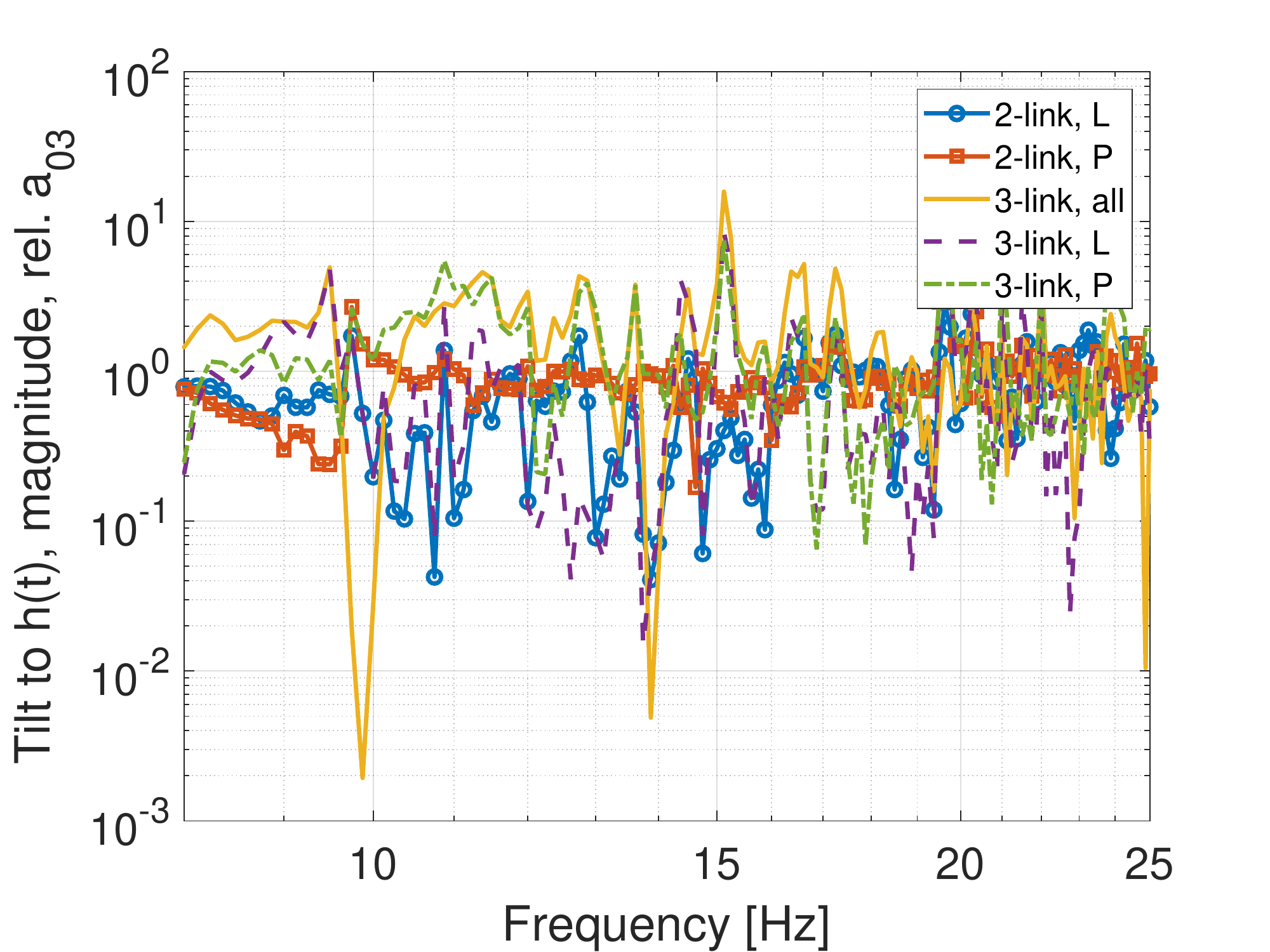}
\caption{Coupling terms used in equations \eqref{eq:ch3} (2-link) and \eqref{eq:ch4} (3-link) considering ITMX SUS L,P. The 2-link curves represent the simple model with either L or P as additional channel. The 3-link curves \{all,L,P\} represent the first, second, and third term on the right-hand-side of equation \eqref{eq:ch4} using L and P as additional channels.}
\label{fig:couplegrav}
\end{figure}
This leaves us with the last step to apply the formalism of Section \ref{sec:cause} to infer the direct coupling between ground tilt and $h(t)$ using correlation measurements. Figure \ref{fig:couplegrav} shows the results split into various terms. The blue-circle curve represents equation \eqref{eq:ch3} using only the ITMX L channel (same equation used for the red-square curve with ITMX P). The dotted blue curve '$a_{03}$' corresponds to equation \eqref{eq:ch4} using both channels ITMX L, P. This equation has three terms that are also plotted for comparison (first term: solid yellow; second term: dashed violet; third term: dash-dotted green). The variation among all these curves indicates that the multi-link corrections are important. 

The inferred $a_{03}$ link is the best possible model we can construct for the direct ground tilt-to-$h(t)$ coupling. However, some assumptions underlying the formalism in Section \ref{sec:cause} limit its accuracy:
\begin{itemize}
\item We have seen in Figure \ref{fig:sushoft} that the contributions of ITMY to correlations with the tiltmeter is significant above 12\,Hz, which means that the 3-link model needs to be extended to a 5-link model to be accurate. However, in light of additional limitations of this analysis, we did not attempt to construct a more accurate model, which would be very hard to solve with additional channels (even numerically because of degeneracies between correlation measurements). 
\item The SNRs of the suspension-platform displacement and rotation measurements are only known approximately, and the same is true for the tiltmeter.
\end{itemize}
The accuracy of the presented analysis will improve with future reduction of low-frequency instrument noise in the LIGO detectors.

\section{Conclusions}
\label{sec:conclusions}
We presented a detailed analysis of the ground-to-$h(t)$ coupling at the corner station of the LIGO Hanford detector. A new formalism to infer physical links using correlation measurements was presented and applied to attempt an identification of gravitational coupling. However, certain approximations were necessary in this analysis and limited its accuracy, and it was not possible to claim a detection of Newtonian noise. It is still the best inference of direct ground-to-$h(t)$ coupling provided so far.

A detailed characterization of the seismic field in terms of seismic speeds and propagation directions was carried out showing that only a handful of local seismic sources determine the dominant ground motion in the NN band. These results were later used to average coupling models over wave-vector space providing the best estimate of gravitational coupling between test mass and seismic field at the LIGO Hanford corner station.

A Wiener filter calculated effectively from 42 days of data (selected from a 70-day observation period) was used to obtain a projection of instrument noise correlated with ground motion. Ground motion was observed using a tiltmeter and an array of seismometers. This noise lay well below other O2 instrument noise, but if other instrument noise would reduce and the noise correlated with ground motion remained unchanged, then this noise could be subtracted by Wiener filtering of the seismic data in the future. This is true irrespective of how this noise enters the GW data, i.e., it does not have to be NN.

The Wiener filter itself showed a complicated structure due to the complexity of the seismic field (e.g., its anisotropy), and because the ground-to-$h(t)$ coupling investigated here happened through both of the inner test masses. It was possible to obtain some understanding of the Wiener filter from the seismic-field characterization, but it did not provide any evidence of gravitational coupling, which is also expected due to the anisotropy of the seismic field (in contrast, gravitational coupling of isotropic seismic fields produces well-understood Wiener filters). The Wiener filter grouped the 30 seismometers into 7 distinct sets of sensors, i.e., only 7 different filters were applied to seismometers, which means that many sensors share the same filter forming ''super-sensors''. To some extent, the creation of super-sensors can be related to sub-optimal placement of seismometers, but the dominant cause is likely that more seismometers than necessary for Wiener filtering were deployed. However, these findings might change in the future since the dominant coupling mechanism during O2 was not necessarily gravitational.

In summary, it was not possible to identify gravitational coupling between seismic fields and test masses. It seems plausible that the dominant ground-correlated noise in $h(t)$ during O2 was due to ground-tilt coupling into the suspension-platform motion since the links investigated in this paper that go through the suspension system have similar strength than the total observed ground-to-$h(t)$ coupling. Still, also the predicted gravitational coupling is similar to the observed coupling at least at certain frequencies of the NN band, which means that new Wiener-filter analyses with improved detector sensitivity will likely yield first observational constraints and tests of NN models.

\acknowledgments
We thank R Schofield for his invaluable comments to this paper. MPR and KV were supported by funding from the NSF under Awards PHY-1607385, PHY1607391, PHY-1912380 and PHY-1912514. MC, JD, and SD are members of the LIGO Laboratory, supported by funding from the U. S. National Science Foundation. LIGO was constructed by the California Institute of Technology and Massachusetts Institute of Technology with funding from the National Science Foundation and operates under cooperative agreement PHY0757058. MC has been supported by NSF grant PHY-1505373 and by the David and Ellen Lee Postdoctoral Fellowship at the California Institute of Technology. BJJS has been supported by the ARC Future Fellowship FT130100329. The authors also gratefully acknowledge the support of the Australian Research Council under the ARC Centre of Excellence for Gravitational Wave Discovery, Grant No. CE170100004 and Linkage Infrastructure, Equipment and Facilities Grant No. LE130100032. The authors acknowledge the use of Matlab and Mathematica for some of the theoretical and numerical analyses.

\raggedright
\bibliographystyle{apsrev}
\bibliography{references}

\end{document}